\documentclass[universe,article,accept,pdftex,oneauthor]{mdpi}

\firstpage{1} 
\pubvolume{1}
\issuenum{1}
\articlenumber{0}
\pubyear{2024}
\copyrightyear{2024}
\externaleditor{Academic Editor: Firstname \linebreak  Lastname}
\datereceived{30 April 2024} 
\daterevised{7 June 2024} 
\dateaccepted{ } 
\datepublished{ } 
\hreflink{https://doi.org/} 

\Title{Galaxy Group Ellipticity Confirms a Younger Cosmos}

\TitleCitation{Galaxy Group Ellipticity Confirms a Younger Cosmos}

\Author{{Yu Rong $^{1, 2}$} 
 \orcidA{}}

\AuthorNames{Yu Rong }

\AuthorCitation{{Rong, Y.} 
}

\address[1,2]{%
 $^{1}$ Department of Astronomy, CAS Key Laboratory for Research in Galaxies and Cosmology, Uni- versity of Science and Technology of China, Hefei, {230052}
, China; rongyua@ustc.edu.cn\\
$^{2}$ School of Astronomy and Space Science, University of Science and Technology of China, Hefei, {230052}, China.}

\abstract{\textls[-15]{We present an analysis of the ellipticities of galaxy groups, derived from the spatial distribution of member galaxies, revealing a notable incongruity between the observed local galaxy groups and their counterparts in the Lambda cold dark matter cosmology. Specifically, our investigation reveals a substantial disparity in the ellipticities of observed groups with masses \mbox{$10^{13.0}<M_{\rm{h}}<10^{14.5}\ {\rm M_{\odot}}\ h^{-1}$} exhibiting significantly higher ellipticities (at a confidence level of approximately $4\sigma$) compared to their simulated counterparts. Notably, the consistent use of the same group finder for identifying galaxy groups in both observational and simulated datasets underscores the robustness of this result. This observation may imply a potential incongruence between the inferred age of the Universe from observations and the predictions of the model, which aligns with the younger Universe hypothesis suggested by the elevated fraction of observed satellite pairs with correlated line-of-sight relative velocities compared to simulations. Our findings significantly strengthen the plausibility of a younger age for our Universe.}}


\begin{document}

\section{Introduction} \label{sec:intro}

Over the past several decades, the~Lambda cold dark matter ($\Lambda$CDM) paradigm has emerged as the predominant model for structure formation in our Universe. Given the initial amplitude of perturbations, structures form hierarchically within the $\Lambda$CDM cosmology, through gravitational collapse and mergers of smaller objects. The~collapse amplifies the initial anisotropy of matter distribution, giving rise to a complex pattern of sheets, filaments, and~knots, delineating large under-dense voids e.g., {\citep{Doroshkevich80,Klypin83,Davis85,Gramann93,Sheth04}.} 
 Galaxies adhere to this network, undergoing a sequence of clustering, interaction, and~merging, gravitating towards denser regions along  well-defined paths, from~the voids to the sheets, the~sheets feeding the filaments, and~the filaments ultimately channeling galaxies towards the knots, culminating in the formation of galaxy groups \citep{Ostriker96,Tempel14,Tempel15,Rong24,Cautun14,Tully14,Karachentsev14}. Consequently, the~motions of member galaxies in galaxy groups encapsulate valuable information about the growth of host structures e.g., \citep{Gu24}. Furthermore, the~distribution and alignment of member galaxies also encode essential information about their assembly history and the dynamical states of their host large-scale structures \citep{Hopkins05,Allgood06,Rong15a,Rong15b,Rong16,Rong19,Rong20}.

Therefore, the~investigation of the motions and distributions of member galaxies in galaxy groups holds the potential to provide insights into the understanding of structure formation and the testing of the $\Lambda$CDM model. In~the local group, the~preferential co-planar and co-rotating arrangement of satellite dwarf galaxies around the Milky Way and M31 suggests a possible origin related to galaxy interactions \citep{Hammer13,Smith16,Banik22}, or~an isotropic accretion of satellite galaxies along cosmic web filaments \citep{Libeskind05,Libeskind14,Buck15,Shao18}. The~spatial distribution of member galaxies in loose galaxy groups such as galaxy pairs and triplets has been observed to align with large-scale filamentary structures, indicating a tendency for clustering along filaments \citep{Tempel15,Rong24}, while the absence of alignment in compact pairs and triplets implies multiple interactions and a potentially equilibrium state within the systems \citep{Trofimov95}. Similarly, an~anisotropic distribution of galaxies in a galaxy cluster is often accompanied by structural interaction, clumpy hot gas distribution, and~non-virilized velocity \mbox{distribution \citep{Weibmann13,Mann12,Maughan08,Hashimoto07,Bauer05,Sereno12,Burke15,Postman12};} in contrast, a~more relaxed galaxy group is expected to exhibit a more isotropic distribution of galaxies and radial alignment of \mbox{satellites \citep{Hopkins05,Allgood06,Rong15a,Rong15b},} as~violent relaxation misanthropizes objects through the time-dependent gravitational potential. Furthermore, the~satellite pairs on both sides of the central galaxies in galaxy groups have been found to display correlated line-of-sight velocities relative to the central galaxies, with~their relative velocities tending to be either positive or negative, indicating a history of infalling for the member galaxies. Drawing on these insights, a~comparison of the distribution and motions of member galaxies in observations and simulations can aid astronomers in testing the {$\Lambda$CDM} 
 cosmology and in evaluating the age and state of the large-scale structure in which the current member galaxies~reside.

In a recent study, Gu~et~al. (2024) \citep{Gu24}  made a noteworthy discovery, revealing that the proportion of satellite pairs exhibiting correlated velocities in observational data significantly exceed that predicted by simulations based on the $\Lambda$CDM framework. This intriguing finding raises the possibility that our Universe may be younger than anticipated or,~alternatively, that the $\Lambda$CDM model may not be as robust as previously thought. Motivated by the implications of this study, we endeavour to explore the distributions of member galaxies within galaxy groups and group ellipticities in the present work. Given that both the analyses of member galaxy motions and distributions within groups are minimally contingent on galaxy formation models and intrinsic properties, it is anticipated that observed galaxy groups would manifest greater ellipticities in comparison to simulations if our Universe is indeed younger. This expectation arises from the premise that a younger Universe corresponds to a shorter duration since the accretion of galaxies into the galaxy groups, thereby rendering the groups less dynamically~relaxed.

In Section~\ref{sec2}, we will provide an introduction to the galaxy group catalogues utilized in both observational and simulated analyses, obtained using the same group finder. Section~\ref{sec3} will encompass a comparative analysis of the ellipticities of groups in observational and simulated datasets. Our findings will be discussed and summarized in Section~\ref{sec4}. In~this paper, we use ``$\log$'' to represent ``$\log_{10}$‘’, and~use $h=H_0/100$ to denote the Hubble~constant.

\section{Galaxy Group Samples in Observation and~Simulation}
\label{sec2}

In our observational analysis, we utilize the galaxy group catalogue developed by Yang~et~al. (2007) \citep{Yang07}, which is derived from the Sloan Digital Sky Survey (SDSS) DR7 data \citep{Abazajian09} and selected using a ``halo-based'' group finder. {{The galaxy} 
 sample utilized for identifying galaxy groups is selected from the New York
University Value Added Galaxy Catalogue (NYU-VAGC) \citep{Blanton05} of the SDSS DR7,  with~redshifts in the range of $0.01< z< 0.20$, spectroscopic completeness
greater than 0.7, and~$r$-band flux-limited of $r<17.77$~mag, where $r$ denotes the extinction-corrected Petrosian magnitude in the $r$-band. The~stellar mass of each galaxy is estimated from the $r$-band absolute magnitude and colour between the $g$ and $r$ bands, $g-r$, by~applying the mass-to-light ratio of Bell~et~al. (2003) \citep{Bell03}. When identifying galaxy groups, the~halo-based group finding algorithm developed by Yang~et~al. (2007) \citep{Yang07} is implemented, which identifies groups based on dark matter halo properties, such as mass and velocity dispersion, expected from the CDM cosmogony. We refer the reader to Yang~et~al. (2005) \citep{Yang05} and Yang~et~al. (2007) \citep{Yang07} for details. This particular group catalogue has demonstrated its efficacy in the examination of large-scale structures and the properties of member galaxies e.g., \citep{Zhang24,Wang23,Shi18,Wang18,Argudo-Fernandez16}. {{In order to reduce} boundary effects, in~this study, we exclude the galaxy groups with $0 < f_{\rm{edge}} < 0.7$, where $f_{\rm{edge}}$ is the fraction of the volume of a group that lies within the survey boundary.}

For comparative purposes, we turn to the galaxy groups within the cosmological \mbox{L-GALAXIES} simulation and conduct a comparison of their ellipticities with those observed. L-GALAXIES is chosen due to its expansive simulation box, spanning approximately $500$~Mpc. The~galaxy formation model, as~detailed in Henriques~et~al. (2015) \citep{Henriques15}, implemented in L-GALAXIES represents an updated iteration of the Munich semi-analytic model e.g., \citep{Croton06,Guo11} established upon the Millennium Simulation \citep{Springel05}. L-GALAXIES employs a Markov Chain Monte Carlo (MCMC) method to explore the high-dimensional parameter space, aiming to replicate the observed galaxy abundance and properties as a function of stellar mass from redshift z$\sim$3 to z$\sim$0. 

Within L-GALAXIES, halos, or~galaxy groups, are identified using the friends-of-friends (FOF) group algorithm, which links particles separated by 0.2 times the mean inter-particle separation \citep{Davis85} at each redshift snapshot. The~SUBFIND algorithm \citep{Springel01} is subsequently utilized to identify the self-bound subhalos within each halo. It is important to note that due to the disparate methods employed for group identification, a~direct comparison of the shapes of the simulated and observed galaxy groups may introduce systematic bias in the estimation of~ellipticity.

In order to facilitate a reliable comparison with the observational data, we constructed a mock catalogue of galaxy groups for L-GALAXIES to account for potential observational selection effects {{by following}
 the method described in Lim~et~al. (2017) \citep{Lim17}. We refer the reader to Lim~et~al. (2017) for details. This involved stacking duplicates of the original simulation box alongside each other to create a suitably large volume. Subsequently, we designated an observer location within the constructed volume, and calculated the redshift and apparent magnitude for each galaxy based on its luminosity, distance, and~relative motion with respect to the observer. Finally, we selected a flux-limited sample ($r<17.77$~mag) of galaxies from a light cone covering the redshift range $0.01 < z < 0.2$, similar to the ranges covered by the SDSS data, and~applied the same group finder used in the study by Yang~et~al. (2007) \citep{Yang07} to identify galaxy groups.}}

\section{Ellipticity of Galaxy~Groups}
\label{sec3}

In this investigation, we leverage the spatial distribution of member galaxies to infer the ellipticities of galaxy groups in both observational and simulated datasets. To~mitigate the influence of redshift-induced distortions, we calculate the ellipticity in the projected sky plane. The~configuration of a galaxy group comprising $N$ member galaxies is approximated using the inertia tensor,
\begin{linenomath}
\begin{equation}
    I_{\alpha\beta}=\Sigma_{i=1}^{N} w_ix_{i,\alpha}x_{i,\beta},
\end{equation}
\end{linenomath}
where $w_i$ denotes the weight assigned to the $i$-th member galaxy, $\alpha$ and $\beta$ are the inertia tensor indices taking values of 1 or 2, and~$x_{i,\alpha}$ represents the position of the $i$-th galaxy relative to the centroid of the galaxy group in the projected sky plane (for this study, we adopt the luminosity-weighted group center). Subsequently, the~axis lengths $a$ and $b$ ($a\geq b$) of the ellipsoidal group are derived from the eigenvalues $\lambda_1$ and $\lambda_2$ ($\lambda_1\geq \lambda_2$) of the inertia tensor, where $a=\sqrt{\lambda_1}$ and $b=\sqrt{\lambda_2}$. 

In this investigation, we employ three distinct methods to estimate the ellipticity of galaxy groups. Method~(1), akin to the approach used by \mbox{Wang~et~al. (2008) \citep{Wang08},} involves setting $w_i=1$ for all member galaxies, signifying equal weights for different member galaxies within a group. Method~(2), akin to the approach proposed by \mbox{Wang~et~al. (2020) \citep{Wang20},} entails setting $w_i=1/R_i^2$, where $R_i$ represents the distance from the $i$-th member galaxy to the group center. In~this method, greater weight is assigned to members in closer proximity to the group center. Method~(3), similar to the approach adopted by Shao~et~al. (2016) \citep{Shao16}, involves setting $w_i= m_{\star,i}$, where $m_{\star,i}$ denotes the stellar mass of the $i$-th member galaxy, thereby assigning greater weight to more massive member~galaxies.

\begin{figure}[H]
 	 \includegraphics[width=0.8\columnwidth]{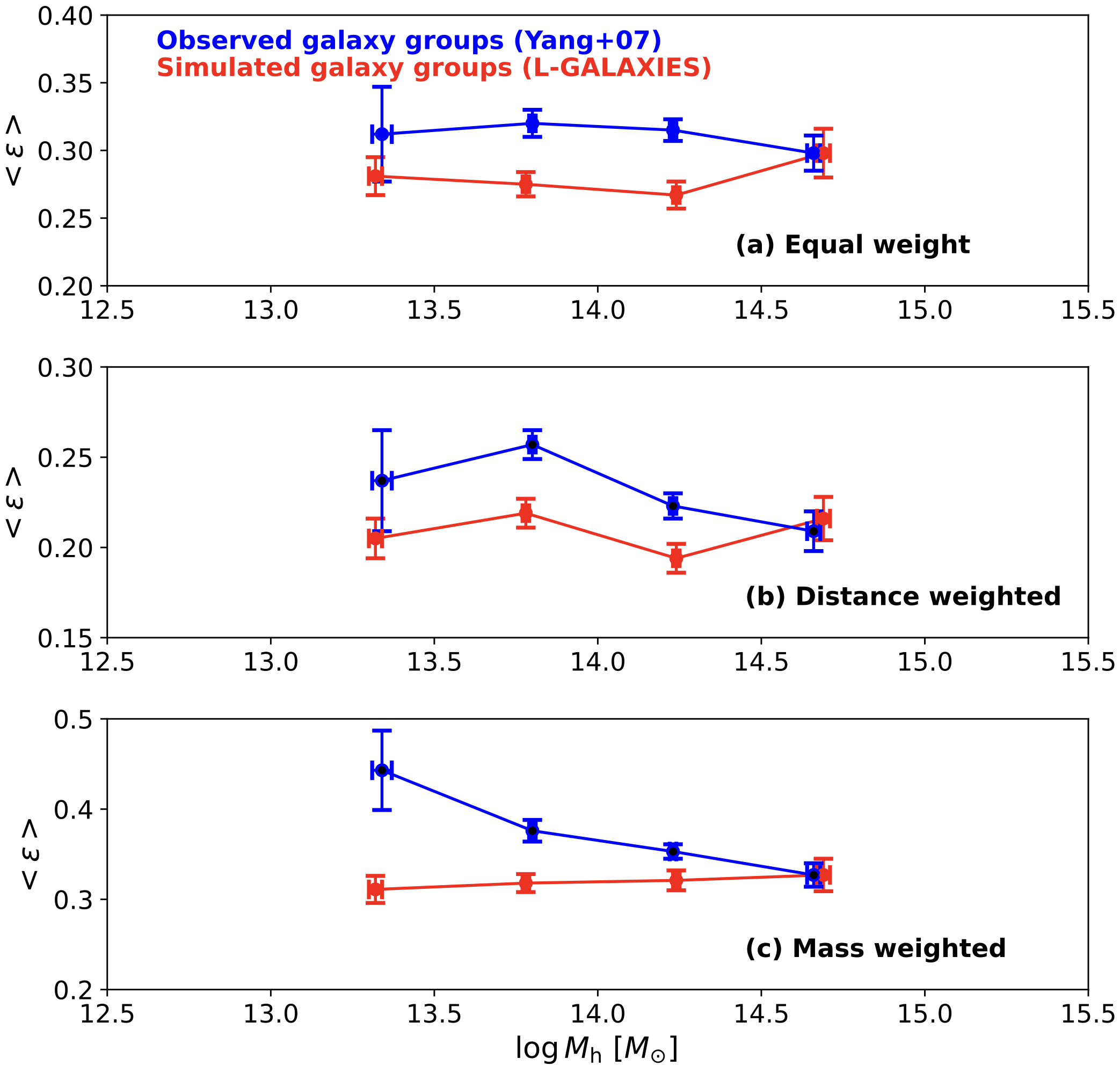}
 	 \caption{\textls[-4]{{The comparison} 
 between average ellipticities of the observed (depicted in blue colour) and simulated  (depicted in red colour)  galaxy groups with the same group finder. Panels~(\textbf{a}--\textbf{c}) correspond to  Method~(1), (2), and~(3), respectively. The~galaxy group samples are split into the different mass bins, within \mbox{$10^{13.0}<M_{\rm{h}}<10^{13.5}\ {\rm M_{\odot}}\ h^{-1}$, $10^{13.5}<M_{\rm{h}}<10^{14.0}\ {\rm M_{\odot}}\ h^{-1}$, $10^{14.0}<M_{\rm{h}}<10^{14.5}\ {\rm M_{\odot}}\ h^{-1}$}, and~$M_{\rm{h}}>10^{14.5}\ {\rm M_{\odot}}\ h^{-1}$. Each point and the corresponding error bar represent the average ellipticity and error of the average value.}
         }\label{ell_mass}
\end{figure}

To ensure accurate estimation of {{the ellipticity} ($\epsilon=1-b/a$) of a galaxy group}, we exclude groups with low richness ($N<20$) or low masses ($M_{\rm{h}}<10^{13.0}\ {\rm M_{\odot}}\ h^{-1}$). Subsequently, we partition the groups in both the simulation and observation datasets into distinct mass bins and compute the average ellipticity, 
 $\langle \epsilon \rangle$, of~groups within each mass bin, along with the error of the average axis ratio (i.e., $\sigma_{\epsilon}/\sqrt{K}$, where $\sigma_{\epsilon}$ represents the standard deviation of axis ratios of groups within a mass bin, and~$K$ denotes the number of groups in the corresponding bin). The~average group axis ratio as a function of group mass for the simulated and observed galaxy groups is depicted in Figure~\ref{ell_mass}.

Our findings indicate that across all mass bins where $M_{\rm{h}} < 10^{14.5}\ {\rm M_{\odot}}\ h^{-1}$, the~observed galaxy groups exhibit a higher average ellipticity compared to the simulated groups, irrespective of the method employed for ellipticity estimation. Specifically, the~observed groups with $M_{\rm{h}} < 10^{14.5}\ {\rm M_{\odot}}\ h^{-1}$ are found to be more elliptical than the simulated counterparts at confidence levels
of $4.7 \sigma$, $3.8 \sigma$, and~$4.2\sigma$ for estimation methods (1), (2), and~(3), respectively. Notably, the~disparity in ellipticity between the simulation and observation appears to diminish as the group mass exceeds $M_{\rm{h}} > 10^{14.5}\ {\rm M_{\odot}}\ h^{-1}$.
\vspace{-6pt}{}
\unskip

\section{Discussion}\label{sec4}

We find that the observed galaxy groups with masses of $M_{\rm{h}} < 10^{14.5}\ {\rm M_{\odot}}\ h^{-1}$ exhibit a statistically significantly higher degree of ellipticity compared to their simulated counterparts. Notably, to~ensure robustness and reliability, we have employed three distinct weighting methodologies for estimating group ellipticities and,~strikingly, our findings remain consistent and independent of the chosen estimation approach. The~discernible disparity in ellipticities between observed galaxy groups and simulation outcomes may indicate that the dynamical states of the observed galaxy groups are potentially younger than anticipated. Notably, the~most massive groups in both simulations and observations, characterized by masses exceeding $M_{\rm{h}} >10^{14.5}\ {\rm M_{\odot}}\ h^{-1}$, exhibit minimal differences in ellipticity, as~evidenced by their closely aligned average ellipticity values (differing at a confidence level of less than $1\sigma$). This convergence in ellipticity could plausibly be attributed to the {{faster}} relaxation and advanced dynamical age of these massive~groups.

Our findings align with the notion of a Universe that is younger than originally postulated, a~conclusion that resonates with the findings of Gu~et~al. (2024) \citep{Gu24} regarding the heightened fractions of satellite pairs in galaxy groups displaying correlated line-of-sight velocities relative to the central galaxies. Importantly, both our investigations and the aforementioned study by Gu~et~al. (2024) \citep{Gu24} draw upon {$N$}
-body cosmological simulations, which are minimally contingent on galaxy formation models and chemical enrichment models. Consequently, the~convergence of our conclusions significantly strengthens the plausibility of a younger age for our~Universe.

On the contrary, it is important to consider that a warm dark matter (WDM) cosmology also has the capacity to engender heightened infalling velocities of member galaxies, which in turn correspond to an elevated proportion of satellites exhibiting substantial and co-oriented velocities relative to the central galaxy \citep{Knebe08} ({{we refer}
 the reader to Figure~10 and Section~3.6 of \citep{Knebe08}}), as~compared to the predictions of the $\Lambda$CDM cosmology. It is worth noting that a WDM cosmology, with~an appropriate particle mass, could feasibly account for the observations reported by Gu~et~al. (2024) \citep{Gu24}. Nevertheless, it is {{also worth}} to acknowledge that a WDM cosmology would concurrently yield a more isotropic distribution of member galaxies within a galaxy group ({{we refer} the reader to Figure~6 \mbox{of \citep{Knebe08}}}), a~trend that contradicts our own findings. Consequently, it is evident that WDM cosmology does not present a viable framework to reconcile these disparate~results.

It is also worth considering that the lower density fluctuations, denoted as $\sigma_8$ in~the early Universe, may also account for the  more elliptical galaxy groups detected in observations \citep{Allgood06}. Particularly, the~$\sigma_8$ of L-GALAXIES is high, with~$\sigma_8\sim 0.9$, whereas the Planck measurements of the cosmic microwave background give $\sigma_8\sim0.8$ \citep{Planck20,Planck16}. Therefore, the~ellipticity difference between the observed and simulated galaxy groups may be explained as the different $\sigma_8$. However, the~lower $\sigma_8$ cannot account for the results of Gu~et~al. (2024) \citep{Gu24} (in \citep{Gu24}, the~fractions of satellite pairs with correlated velocities in \mbox{TNG300 \citep{Nelson18,Nelson19},} which applies $\sigma_8\sim 0.8$, are still much lower than the observables). Consequently, it is evident that the lower $\sigma_8$ also does not present a viable framework to reconcile these disparate~results.

We also acknowledge the potential influence of background or foreground interlopers, introduced by the group finder, which could conceivably contribute to a more spheroidal group shape. However, it is crucial to emphasize that in this study, the~identical group finder was uniformly applied to identify both the observed and simulated groups. Consequently, the~level of contamination in both the observational and simulated datasets should theoretically be comparable, rendering it implausible for this factor alone to account for such a substantial~disparity. 

Furthermore, we systematically varied the threshold for selecting groups, transitioning from a richness criterion of $N>20$ to $N>10$, or~$N>30$, among~others, and~consistently arrived at a similar overarching conclusion. Additionally, we conducted an analysis utilizing the brightest galaxies or most massive galaxies as the group centres and,~remarkably, the~adjustment yielded no discernible changes in ellipticity~difference.

{{Finally, it is worth }
 noting that the recent observational research on globular clusters by Llorente de Andr\'es (2024) \citep{Llorente24} suggests that the age of the Universe may be older than theorized. Furthermore, observations with the James Webb Space Telescope (JWST) have also revealed the presence of numerous high-redshift massive galaxies, e.g., \citep{Boyett24,Haslbauer22}. Given that galaxies require sufficient time for star formation, these findings may also indicate that the Universe is somewhat older than expected. However, it should be pointed out that these observational studies heavily depend on the assumed galaxy formation models and chemical evolution models, which are quite uncertain, particularly for the cases in the early Universe. In~contrast, our work and that of Gu~et~al. (2024) \citep{Gu24} are primarily based on the hierarchical structure formation theory within the framework of $\Lambda$CDM, with~little relevance to galaxy formation models. Therefore, our work and that of Gu~et~al. (2024) may be more credible.}


\section{Conclusions}\label{sec5}

Taking advantage of the galaxy group catalogue derived from observational data in conjunction with the mock group catalogue generated from cosmological simulations, both obtained through the application of a consistent group finder, our analysis reveals that observed galaxy groups with masses of $M_{\rm{h}} < 10^{14.5}\ {\rm M_{\odot}}\ h^{-1}$ exhibit a statistically significantly higher degree of ellipticity compared to their simulated counterparts at~a confidence level of approximately $4\sigma$.

{{We find that the discernible }
  disparity in ellipticities between the observed galaxy groups and simulation outcomes cannot be explained by a warm dark matter cosmology or lower density fluctuations in the early Universe, but~suggest that the observed galaxy groups are potentially younger than anticipated and,~thus, that the observed cosmic age may be younger than the theoretical prediction.}

{{Our conclusions are consistent } 
with those of Gu~et~al. (2024), further confirming that the actual age of the Universe may be younger than expected based on theory.}

\vspace{6pt}


\funding{Y.R. acknowledges supports from the NSFC grant 12273037, and~the CAS Pioneer Hundred Talents Program (Category B), as~well as the USTC Research Funds of the Double First-Class Initiative (grant No.~YD2030002013).}

\dataavailability{Data available if requested.} 




\acknowledgments{Y. R. thanks Huiyuan Wang and Enci Wang in USTC, as~well as Yougang Wang in NAOC, for~helpful discussion on ellipticity~estimation.}

\conflictsofinterest{The author declares no conflicts of~interest.} 



\abbreviations{Abbreviations}{
The following abbreviations are used in this manuscript:\\

\noindent 
\begin{tabular}{@{}ll}
$\Lambda$CDM & Lambda cold dark matter\\
FOF & Friends-of-friends\\
MCMC & Markov Chain Monte Carlo\\
WDM & Warm dark matter\\
JWST & James Webb Space Telescope \\
SDSS & loan Digital Sky Survey 
\end{tabular}
}


\begin{adjustwidth}{-\extralength}{0cm}

\reftitle{References}



\begin{thebibliography}{999}
	
\bibitem[\protect\citeauthoryear{Doroshkevich et al.}{1980}]{Doroshkevich80} Doroshkevich, A.G.; Kotok, E.V.; {Polyudov, A.N.; Shandarin, S.F.; Sigov, Y.S.;} Novikov, I.D. Two-dimensional simulation of the gravitational system dynamics and formation of the large-scale structure of the universe.  \emph{Mon. Not. R. Astron. Soc.}  \textbf{1980}, \emph{192}, 321.	
\bibitem[\protect\citeauthoryear{Klypin \& Shandarin}{1983}]{Klypin83} Klypin, A.A.;  Shandarin, S.F. Three-dimensional numerical model of the formation of large-scale structure in the Universe. \emph{Mon. Not. R. Astron. Soc.}  \textbf{1983}, \emph{204}, 891.	
\bibitem[\protect\citeauthoryear{Davis et al.}{1985}]{Davis85} Davis, M.; Efstathiou, G.; Frenk, C.S.; White, S.D.M. The evolution of large-scale structure in a universe dominated by cold dark matter. \emph{Astrophys. J.}  \textbf{1985}, \emph{292}, 371.
\bibitem[\protect\citeauthoryear{Gramann}{1993}]{Gramann93} Gramann, M. \emph{Astrophys. J.} An Improved Reconstruction Method for Cosmological Density Fields.  \textbf{1993}, \emph{405}, 449.
\bibitem[\protect\citeauthoryear{Sheth \& van de Weygaert}{2004}]{Sheth04} Sheth, R.K.; van de Weygaert, R. A hierarchy of voids: much ado about nothing. \emph{Mon. Not. R. Astron. Soc.}  \textbf{2004}, \emph{350}, 517.
\bibitem[\protect\citeauthoryear{Ostriker \& Cen}{1996}]{Ostriker96} Ostriker, J.P.; Cen, R. Hydrodynamic Simulations of the Growth of Cosmological Structure: Summary and Comparisons among Scenarios. \emph{Astrophys. J.}  \textbf{1996}, \emph{464}, 27.
\bibitem[\protect\citeauthoryear{Tempel}{2014}]{Tempel14} Tempel, E. Cosmology: Meet the Laniakea supercluster. \emph{Nature} \textbf{2014}, \emph{513}, 41.
\bibitem[\protect\citeauthoryear{Tempel \& Tamm}{2015}]{Tempel15} Tempel, E.; Tamm, A. Galaxy pairs align with Galactic filaments. \emph{A\&A} \textbf{2015}, \emph{576L}, 5.
\bibitem[\protect\citeauthoryear{Rong et al.}{2024}]{Rong24} Rong, Y.; Shen, J.; Hua, Z.  Galaxy triplets alignment in large-scale filaments. \emph{Mon. Not. R. Astron. Soc.}  \textbf{2024}, \emph{531L}, 9.
\bibitem[\protect\citeauthoryear{Cautun et al.}{2014}]{Cautun14} Cautun, M.; van de Weygaert, R.; Jones, B.J.T.; Frenk, C.S. Evolution of the cosmic web. \emph{Mon. Not. R. Astron. Soc.}   \textbf{2014}, \emph{441}, 2923.
\bibitem[\protect\citeauthoryear{Tully et al.}{2014}]{Tully14} Tully, R.B.; Courtois, H.; Hoffman, Y.; Pomar\`ede, D. The Laniakea supercluster of galaxies. \emph{Nature} \textbf{2014},  \emph{513}, 71.
\bibitem[\protect\citeauthoryear{Karachentsev et al.}{2014}]{Karachentsev14} Karachentsev, I.D.; Karachentseva, V.E.; Nasonova, O.G. Motions of Galaxies in the Bootes Strip. \emph{Astrophysics} \textbf{2014}, \emph{57}, 457.
\bibitem[\protect\citeauthoryear{Gu et al.}{2024}]{Gu24} Gu, Q.; Guo, Q.; Cautun, M.; Shao, S.; Pei, W.; Wang, W.; Gao, L.; Wang, J. 
 A younger Universe implied by satellite pair correlations from SDSS observations of massive galaxy groups. \emph{Nat. Astron.} \textbf{2024}, \emph{8}, 538.
\bibitem[\protect\citeauthoryear{Hopkins et al.}{2005}]{Hopkins05} Hopkins, P.F.; Bahcall, N.A.; Bode, P.  Cluster Alignments and Ellipticities in $\Lambda$CDM Cosmology. \emph{Astrophys. J.}  \textbf{2005}, \emph{618}, 1.
\bibitem[\protect\citeauthoryear{Allgood et al.}{2006}]{Allgood06} Allgood, B., Flores, R. A., Primack, J. R., Kravtsov, A.V.; Wechsler, R.H.; Faltenbacher, A.; Bullock, J.S. The shape of dark matter haloes: dependence on mass, redshift, radius and formation. \emph{Mon. Not. R. Astron. Soc.}  \textbf{2006}, \emph{367}, 1781.
\bibitem[\protect\citeauthoryear{Rong et al.}{2015a}]{Rong15a} Rong, Y.; Yi, S.-X.; Zhang, S.-N.; Tu, H. Radial alignment of elliptical galaxies by the tidal force of a cluster of galaxies. \emph{Mon. Not. R. Astron. Soc.}  \textbf{2015}, \emph{451}, 2536.
\bibitem[\protect\citeauthoryear{Rong et al.}{2015b}]{Rong15b} Rong, Y.; Zhang, S.-N.; Liao, J.-Y. Primordial alignment of elliptical galaxies in intermediate redshift clusters. \emph{Mon. Not. R. Astron. Soc.}  \textbf{2015}, \emph{453}, 1577.
\bibitem[\protect\citeauthoryear{Rong et al.}{2016}]{Rong16} Rong, Y.; Zhang, S.-N.; Liao, J.-Y. Galaxy alignment as a probe of large-scale filaments. \emph{Mon. Not. R. Astron. Soc.}  \textbf{2016}, \emph{455}, 2267.
\bibitem[\protect\citeauthoryear{Rong et al.}{2019}]{Rong19} Rong, Y.; Zhang, S.-N.; Liao, J.-Y. The Next Generation Fornax Survey (NGFS). VI. The Alignment of Dwarf Galaxies in the Fornax Cluster. \emph{Astrophys. J.}  \textbf{2019}, \emph{883}, 56.
\bibitem[\protect\citeauthoryear{Rong et al.}{2020}]{Rong20} Rong, Y.; Zhang, S.-N.; Liao, J.-Y. Exploring the origin of ultra-diffuse galaxies in clusters from their primordial alignment. \emph{Mon. Not. R. Astron. Soc.}  \textbf{2020}, \emph{498L}, 72.
\bibitem[\protect\citeauthoryear{Hammer et al.}{2013}]{Hammer13} Hammer, F.; {Yang, Y.; Fouquet, S.; Pawlowski, M.S.; Kroupa, P.; Puech, M.; Flores, H.; Wang, J.} The vast thin plane of M31 corotating dwarfs: an additional fossil signature of the M31 merger and of its considerable impact in the whole Local Group. \emph{Mon. Not. R. Astron. Soc.}   \textbf{2013}, \mbox{\emph{431}, 3543.}
\bibitem[\protect\citeauthoryear{Smith et al.}{2016}]{Smith16} Smith, R.; Duc, P.A.; Bournaud, F.; Yi, S.K. A Formation Scenario for the Disk of Satellites: Accretion of Satellites during Mergers. \emph{Astrophys. J.}  \textbf{2016}, \emph{818}, 11.
\bibitem[\protect\citeauthoryear{Banik et al.}{2022}]{Banik22} Banik, I.; {Thies, I.; Truelove, R.; C.; lish, G.; Famaey, B.; Pawlowski, M.S.; Ibata, R.; Kroupa, P.} 3D hydrodynamic simulations for the formation of the Local Group satellite planes. \emph{Mon. Not. R. Astron. Soc.}   \textbf{2022}, \emph{513}, 129.
\bibitem[\protect\citeauthoryear{Libeskind et al.}{2005}]{Libeskind05} Libeskind, N.I.; {Frenk, C.S.; Cole, S.; Helly, J.C.; Jenkins, A.; Navarro, J.F.; Power, C}. The distribution of satellite galaxies: the great pancake. \emph{Mon. Not. R. Astron. Soc.}  \textbf{2005}, \emph{363}, 146.
\bibitem[\protect\citeauthoryear{Libeskind et al.}{2014}]{Libeskind14} Libeskind, N.I.; {Knebe, A.; Hoffman, Y.; Gottl\"ober, S.} The universal nature of subhalo accretion. \emph{Mon. Not. R. Astron. Soc.}  \textbf{2014}, \emph{443}, 1274.
\bibitem[\protect\citeauthoryear{Buck et al.}{2015}]{Buck15} Buck, T.; Maccio, A.V.; Dutton, A.A. Evidence for Early Filamentary Accretion from the Andromeda Galaxy's Thin Plane of Satellites. \emph{Astrophys. J.}  \textbf{2015}, \emph{809}, 49.
\bibitem[\protect\citeauthoryear{Shao et al.}{2018}]{Shao18} Shao, S.; {Cautun, M.; Frenk, C.S.; Gr, ; R.J.; Gómez, F.A.; Marinacci, F.; Simpson, C.M.} The multiplicity and anisotropy of galactic satellite accretion. \emph{Mon. Not. R. Astron. Soc.}  \textbf{2018}, \emph{476}, 1796.
\bibitem[\protect\citeauthoryear{Trofimov \& Chernin}{1995}]{Trofimov95} Trofimov, A.V.; Chernin, A.D. Wide triplets of galaxies and the problem of hidden mass. \emph{AZh} \textbf{1995}, \emph{72}, 308.
\bibitem[\protect\citeauthoryear{Wei{\ss}mann et al.}{2013}]{Weibmann13} Wei{\ss}mann, A.; B$\rm{\ddot{o}}$hringer, H.; Chon, G. Probing the evolution of the substructure frequency in galaxy clusters up to $z\sim 1$ \emph{A\&A} \textbf{2013}, \emph{555}, 147.
\bibitem[\protect\citeauthoryear{Mann \& Ebeling}{2012}]{Mann12} Mann, A.W.; Ebeling, H. X-ray-optical classification of cluster mergers and the evolution of the cluster merger fraction. \emph{Mon. Not. R. Astron. Soc.}  \textbf{2012}, \emph{420}, 2120.
\bibitem[\protect\citeauthoryear{Maughan et al.}{2008}]{Maughan08} Maughan, B.J.; Forman, C.J.; Van Speybroeck, L. Images, Structural Properties, and Metal Abundances of Galaxy Clusters Observed with Chandra ACIS-I at $0.1 < z < 1.3$. \emph{Astrophys. J. Suppl. Ser.} \textbf{2008}, \emph{174}, 117.
\bibitem[\protect\citeauthoryear{Hashimoto et al.}{2007}]{Hashimoto07} Hashimoto, Y.; B$\rm{\ddot{o}}$hringer, H.; Henry, J.P.; Hasinger, G.; Szokoly, G. Robust quantitative measures of cluster X-ray morphology, and comparisons between cluster characteristics. \emph{A\&A} \textbf{2007}, \emph{467}, 485.
\bibitem[\protect\citeauthoryear{Bauer et al.}{2005}]{Bauer05} Bauer, F.E.; Fabin, A.C.; S.; Er, J.S.; Allen, S.W.; Johnstone, R.M. The prevalence of cooling cores in clusters of galaxies at $z\sim 0.15\--0.4$. \emph{Mon. Not. R. Astron. Soc.}  \textbf{2005},  \emph{359}, 1481.
\bibitem[\protect\citeauthoryear{Sereno \& Zitrin}{2012}]{Sereno12} Sereno, M.; Zitrin, A. Triaxial strong-lensing analysis of the $z > 0.5$ MACS clusters: the mass-concentration relation. \emph{Mon. Not. R. Astron. Soc.}  \textbf{2012}, \emph{419}, 3280.
\bibitem[\protect\citeauthoryear{Burke et al.}{2015}]{Burke15} Burke, C.; Hilton, M.; Collins, C. Coevolution of brightest cluster galaxies and intracluster light using CLASH. \emph{Mon. Not. R. Astron. Soc.}   \textbf{2015}, \emph{449}, 2353.
\bibitem[\protect\citeauthoryear{Postman et al.}{2012}]{Postman12} Postman, M.; Coe, D.; Benitez, N.; {Bradley, L.; Broadhurst, T.; Donahue, M.; Ford, H.; Graur, O.; Graves, G.; Jouvel, S.};~et~al. The Cluster Lensing and Supernova Survey with Hubble: An Overview. \emph{Astrophys. J. Suppl. Ser.} \textbf{2012}, \emph{199}, 25.
\bibitem[\protect\citeauthoryear{Yang et al.}{2007}]{Yang07} Yang, X.; Mo, H.J.; van den Bosch, F.C.; Pasquali, A.; Li, C.; Barden, M. Galaxy Groups in the SDSS DR4. I. The Catalog and Basic Properties. \emph{Astrophys. J.}  \textbf{2007}, \emph{671}, 153.
\bibitem[\protect\citeauthoryear{Abazajian et al.}{2009}]{Abazajian09}Abazajian, K.N.; {Adelman-McCarthy, J.K.; Agüeros, M.A.; Allam, S.S.; Prieto, C.A.; An D.; Anderson, K.S.; Anderson, S.F.; Annis, J.; Bahcall, N.A.};~{et~al.} The Seventh Data Release of the Sloan Digital Sky Survey. \emph{Astrophys. J. Suppl. Ser.} \textbf{2009}, \emph{182}, 543.
\bibitem[\protect\citeauthoryear{Blanton et al.}{2005}]{Blanton05} Blanton, M.R.; Eisenstein, D.; Hogg, D.W.; Schlegel, D.J.; Brinkmann, J. Relationship between Environment and the Broadband Optical Properties of Galaxies in the Sloan Digital Sky Survey. \emph{Astrophys. J.}   \textbf{2005}, \emph{629}, 143.
\bibitem[\protect\citeauthoryear{Bell et al.}{2003}]{Bell03} Bell, E.F.; McIntosh, D.H.; Katz, N.; Weinberg, M.D. The Optical and Near-Infrared Properties of Galaxies. I. Luminosity and Stellar Mass Functions. \emph{Astrophys. J. Suppl. Ser.}  \textbf{2003}, \emph{149}, 289.
\bibitem[\protect\citeauthoryear{Yang et al.}{2005}]{Yang05} Yang, X.; Mo, H.J.; van den Bosch, F.C.; Jing, Y.P.  \emph{Mon. Not. R. Astron. Soc.}  \textbf{2005}, \emph{356}, 1293.
\bibitem[\protect\citeauthoryear{Zhang et al.}{2024}]{Zhang24} Zhang, Z.; Wang, H.; Luo, W.; Mo, H.; Zhang, J.; Yang, X.; Li, H.; Li, Q. A halo-based galaxy group finder: calibration and application to the 2dFGRS. Halo Mass-observable Proxy Scaling Relations and Their Dependencies on Galaxy and Group Properties. \emph{Astrophys. J.}  \textbf{2024}, \emph{960}, 71.
\bibitem[\protect\citeauthoryear{Wang et al.}{2023}]{Wang23} Wang, K.; Peng, Y.; Chen, Y.  Dissect two-halo galactic conformity effect for central galaxies: the dependence of star formation activities on the large-scale environment. \emph{Mon. Not. R. Astron. Soc.}  \textbf{2023}, \emph{523}, 1268.
\bibitem[\protect\citeauthoryear{Shi et al.}{2018}]{Shi18} Shi, F.; {Yang, X.; Wang, H.; Zhang, Y.; Mo, H.J.; van den Bosch, F.C.; Luo, W.; Tweed, D.; Li, S.; Liu, C.};~et~al. Mapping the Real Space Distributions of Galaxies in SDSS DR7. II. Measuring the Growth Rate, Clustering Amplitude of Matter, and Biases of Galaxies at Redshift 0.1. \emph{Astrophys. J.}  \textbf{2018}, \emph{861}, 137.
\bibitem[\protect\citeauthoryear{Wang et al.}{2018}]{Wang18} Wang, E.; Wang, H.; Mo, H.; van den Bosch, F.C.; Lim, S.H.; Wang, L.; Yang, X.; Chen, S. The Dearth of Differences between Central and Satellite Galaxies. II. Comparison of Observations with L-GALAXIES and EAGLE in Star Formation Quenching. \emph{Astrophys. J.}  \textbf{2018}, \emph{864}, 51.
\bibitem[\protect\citeauthoryear{Argudo-Fern\'andez et al.}{2016}]{Argudo-Fernandez16} Argudo-Fern\'andez, M., Shen, S.; Sabater, J.; Duarte, Puertas, S.; Verley, S.; Yang, X. The effect of local and large-scale environments on nuclear activity and star formation. \emph{A\&A}  \textbf{2016}, \emph{592A}, 30.
\bibitem[\protect\citeauthoryear{Henriques et al.}{2015}]{Henriques15} Henriques, B.M.B.; White, S.D.M.; Thomas, P.A.; Angulo, R.; Guo, Q.; Lemson, G.; Springel, V.; Overzier, R. Galaxy formation in the Planck cosmology - I. Matching the observed evolution of star formation rates, colours and stellar masses. \emph{Mon. Not. R. Astron. Soc.}  \textbf{2015}, \emph{451}, 2663.
\bibitem[\protect\citeauthoryear{Croton et al.}{2006}]{Croton06} Croton, D.J.; {Springel, V.; White, S.D.; De Lucia, G.; Frenk, C.S.; Gao, L.; Jenkins, A.; Kauffmann, G.; Navarro, J.F.; Yoshida, N.} The many lives of active galactic nuclei: cooling flows, black holes and the luminosities and colours of galaxies. \emph{Mon. Not. R. Astron. Soc.}  \textbf{2006}, \emph{365}, 11.
\bibitem[\protect\citeauthoryear{Guo et al.}{2011}]{Guo11} Guo, Q.; {White, S.; Boylan-Kolchin, M.; De Lucia, G.; Kauffmann, G.; Lemson, G.; Li, C.; Springel, V.; Weinmann, S.} From dwarf spheroidals to cD galaxies: simulating the galaxy population in a $\Lambda$CDM cosmology. \emph{Mon. Not. R. Astron. Soc.}  \textbf{2011}, \emph{413}, 101.
\bibitem[\protect\citeauthoryear{Springel et al.}{2005}]{Springel05} Springel, V.; {White, S.D.; Jenkins, A.; Frenk, C.S.; Yoshida, N.; Gao, L.; Navarro, J.; Thacker, R.; Croton, D.; Helly, J.};~et~al. Simulations of the formation, evolution and clustering of galaxies and quasars. \emph{Nature} \textbf{2005}, \emph{435}, 629.
\bibitem[\protect\citeauthoryear{Springel et al.}{2001}]{Springel01} Springel, V.; White, S.D.M.; Tormen, G.; Kauffmann, G. Populating a cluster of galaxies - I. Results at $z=0$. \emph{Mon. Not. R. Astron. Soc.}  \textbf{2001}, \emph{328}, 726.
\bibitem[\protect\citeauthoryear{Lim et al.}{2017}]{Lim17} Lim, S.H.; Mo, H.J.; Lu, Y.; Wang, H.; Yang, X. Galaxy groups in the low-redshift Universe. \emph{Mon. Not. R. Astron. Soc.}  \textbf{2017}, \emph{470}, 2982.
\bibitem[\protect\citeauthoryear{Wang et al.}{2008}]{Wang08} Wang, Y.; Yang, X.; Mo, H.J.; Li, C.; van den Bosch, F.C.; Fan, Z.; Chen, X. Probing the intrinsic shape and alignment of dark matter haloes using SDSS galaxy groups. \emph{Mon. Not. R. Astron. Soc.}  \textbf{2008}, \emph{385}, 1511.
\bibitem[\protect\citeauthoryear{Wang et al.}{2020}]{Wang20} Wang, P.; Libeskind, N.I.; Tempel, E.; Pawlowski, M.S.; Kang, X.; Guo, Q. The Alignment of Satellite Systems with Cosmic Filaments in the SDSS DR12. \emph{Astrophys. J.}  \textbf{2020}, \emph{900}, 129.
\bibitem[\protect\citeauthoryear{Shao et al.}{2016}]{Shao16} Shao, S.; {Cautun, M.; Frenk, C.S.; Gao, L.; Crain, R.A.; Schaller, M.; Schaye, J.; Theuns, T.} Alignments between galaxies, satellite systems and haloes. \emph{Mon. Not. R. Astron. Soc.}  \textbf{2016}, \emph{460}, 3772.
\bibitem[\protect\citeauthoryear{Knebe et al.}{2008}]{Knebe08} Knebe, A.; Arnold, B.; Power, C.; Gibson, B.K. The dynamics of subhaloes in warm dark matter models. \emph{Mon. Not. R. Astron. Soc.}  \textbf{2008}, \emph{386}, 1029.
\bibitem[\protect\citeauthoryear{Planck Collaboration}{2020}]{Planck20} Aghanim, N.~et~al. [Planck Collaboration] Planck 2018 results. VI. Cosmological parameters. \emph{A\&A} \textbf{2020}, \emph{641A}, 6.
\bibitem[\protect\citeauthoryear{Planck Collaboration}{2016}]{Planck16} Ade, P.A.~et~al. [Planck Collaboration] Planck 2015 results. XIII. Cosmological parameters. \emph{A\&A} \textbf{2016}, \emph{594A}, 13.
\bibitem{Nelson18} Nelson, D.; {Pillepich, A.; Springel, V.; Weinberger, R.; Hernquist, L.; Pakmor, R.; Genel, S.; Torrey, P.; Vogelsberger, M.; \mbox{Kauffmann, G.}};~et~al. First results from the IllustrisTNG simulations: The galaxy colour bimodality. First results from the IllustrisTNG simulations: the galaxy colour bimodality. \emph{Mon. Not. R. Astron. Soc.}  \textbf{2018}, \emph{475}, 624.
\bibitem{Nelson19} Nelson, D.; {Springel, V.; Pillepich, A.; Rodriguez-Gomez, V.; Torrey, P.; Genel, S.; Vogelsberger, M.; Pakmor, R.; Marinacci, F.; Weinberger, R.};~et~al. {The IllustrisTNG simulations: Public data release.} The IllustrisTNG simulations: public data release. \emph{Comput. Astrophys. Cosmol.} \textbf{2019}, \emph{6}, 2.
\bibitem[\protect\citeauthoryear{Llorente de Andr\'es}{2024}]{Llorente24} Llorente de Andr\'es, F. Some Old Globular Clusters (and Stars) Inferring That the Universe Is Older Than Commonly Accepted. \emph{ 
American Journal of Astronomy and Astrophysics.} \textbf{2024}, \emph{11}, 1
\bibitem[\protect\citeauthoryear{Boyett et al.}{2024}]{Boyett24} Boyett, K.; {Boyett, K.; Trenti, M.; Leethochawalit, N.; Calabró, A.; Metha, B.; Roberts-Borsani, G.; Dalmasso, N.; Yang, L.; Santini, P.; Treu, T.};~et~al. A massive interacting galaxy 510 million years after the Big Bang. \emph{Nat. Astron.} \textbf{2024}, \emph{8}, 657.
\bibitem[\protect\citeauthoryear{Haslbauer et al.}{2022}]{Haslbauer22} Haslbauer, M.; Kroupa, P.; Zonoozi, A.H.; Haghi, H. Has JWST Already Falsified Dark-matter-driven Galaxy Formation? \emph{Astrophys. J.}  \textbf{2022}, \emph{939L}, 31.


\end{thebibliography}


\PublishersNote{}
\end{adjustwidth}
\end{document}